\documentclass[12pt]{article}
\usepackage{graphicx}

\usepackage{times}

\newenvironment{sciabstract}{%
\begin{quote} \bf}
{\end{quote}}

\newcounter{lastnote}

\title{Graphene: A Pseudochiral Fermi Liquid}

\author
{Marco Polini,$^{1*}$ Reza Asgari,$^{2}$ Yafis Barlas,$^{3}$ T. Pereg-Barnea,$^{3}$ A.H. MacDonald$^{3}$\\
\normalsize{$^{1}$NEST-CNR-INFM and Scuola Normale Superiore, I-56126 Pisa, Italy}\\
\normalsize{$^{2}$Institute for Studies in Theoretical Physics and Mathematics, Tehran 19395-5531, Iran}\\
\normalsize{$^{3}$Department of Physics, University of Texas at Austin, Austin TX 78712, USA}\\
\\
\normalsize{$^\ast$To whom correspondence should be addressed; E-mail:  m.polini@sns.it}
}

\date{}

\begin{document}

\baselineskip24pt

\maketitle

\begin{sciabstract}
Doped graphene sheets are pseudochiral two-dimensional Fermi liquids with 
abnormal electron-electron interaction physics. We address graphene's Fermi liquid properties 
quantitatively using a microscopic random-phase-approxima-
tion theory and comment on the importance of using exchange-correlation potentials based 
on the properties of a chiral two-dimensional electron gas in density-functional-theory 
applications to graphene nanostructures.
\end{sciabstract}

\eject

\noindent 
The seminal work of Novoselov {\it et al.} ({\it 1, 2}) and 
Zhang {\it et al.} ({\it 3}), which demonstrated electrical contact to an isolated 
single-layer graphite flake, has heightened interest in the graphene two-dimensional 
electron system (2DES).  Graphene's honeycomb lattice (Fig.~1) has two-atoms per unit 
cell and its $\pi$-valence band and $\pi^*$-conduction band touch at two inequivalent 
points in the honeycomb lattice Brillouin-zone.  The energy bands near $K$ and 
$K'$ are described at low energies by a spin-independent massless Dirac equation ({\it 4}):
\noindent
\begin{equation}\label{eq:Dirac}
{\cal H}= v \tau \left( \sigma_1 \, p_1 + \sigma_2 \, p_2 \right)\,,
\end{equation}
where $\tau = \pm 1$ for $K$ and $K'$ valleys, the $p_i$ are envelope function
momentum operators, and the $\sigma_i$ are Pauli matrices which act on the sublattice pseudospin
degree-of-freedom.  Like the bands of semiconductor heterojunction 2DESs,
graphene's bands can be described by a continuum model with a single parameter, in this 
case the Dirac velocity $v$ instead of the band mass $m_{\rm b}$.  Just as a semiconductor provides a realization 
of non-relativistic quantum mechanics with a material-dependent mass $m_{\rm b}$, graphene provides a realization of 
relativistic quantum mechanics with a material specific velocity of light.
When Coulombic electron-electron interactions are included,
doped graphene represents a new type of many-electron problem,
distinct from both an ordinary 2DES and from quantum electrodynamics.
The Hamiltonian in Eq.~(\ref{eq:Dirac}) differs from a Schr\"odinger equation in two crucial respects: i) its 
spectrum is not bounded from below and ii) its eigenstates have definite projection of pseudospin
along the direction of momentum, {\em i.e.} definite pseudochirality, rather than definite 
pseudospin. We refer to the graphene 2DES as the chiral 2DES (C2DES).
In this article we explain why the C2DES and the ordinary 2DES have distinctly 
different Fermi liquid properties.    

2DESs have been a fertile source ({\it 5}) of surprising new physics for more than four decades, most notably in the 
presence of a strong magnetic field when they exhibit the integer ({\it 6}) 
and fractional ({\it 7}) quantum Hall effects.  Although the exploration of 
graphene is still at an early stage, it is already clear ({\it 1--3, 8--10}) 
that the strong field properties of a C2DES are different from and as rich as those of a 
semiconductor heterojunction 2DES.
In the absence of a field, ordinary 2DESs are normal Fermi liquids ({\it 11, 12}) 
in which interactions alter the Fermi velocity $v \to v^\star$, introduce marginally
irrelevant effective interactions between quasiparticles on the circular Fermi 
surface, and diminish the fraction $Z$ of the spectral weight in the one-particle Green's function
associated with its quasiparticle peak.  The Fermi liquid phenomenologies of a C2DES and an
ordinary 2DES have the same structure, since both systems are isotropic and have a single circular Fermi surface
as illustrated in Fig.~2. 
The strength of interaction effects in an ordinary 2DES increases with decreasing carrier density.
At low densities, the quasiparticle weight $Z$ is small, the velocity is suppressed, the 
charge compressibility changes sign from positive to negative, and the spin-susceptibility 
is strongly enhanced.  These effects, described with reasonable consistency ({\it 13--19})
by theory and experiment, emerge from an interplay between exchange interactions and quantum
fluctuations of charge and spin in the 2DES.  In the C2DES we find that interaction effects 
also strengthen with decreasing density, although more slowly, that the quasiparticle weight 
$Z$ tends to larger values, that the velocity is enhanced rather than suppressed, and that
the influence of interactions on the compressibility and the 
spin-susceptibility changes sign. These 
qualitative differences are due to exchange interactions between electrons near the 
Fermi surface and electrons in the negative energy sea, to quasiparticle chirality, and 
to interband contributions to C2DES charge and spin fluctuations.
The interband excitations are closely 
analogous to virtual particle-antiparticle excitations of a truly relativistic electron gas.

The technical calculation ({\it 20}) on which our conclusions are based is an evaluation of the 
electron self-energy $\Sigma$ of the C2DES near the quasiparticle-pole. 
$\Sigma$ describes the interaction of a single electron near the 2DES Fermi surface with all 
states inside the Fermi sea, and with virtual particle-hole and collective excitations of the entire Fermi sea,
as illustrated in Fig.~2.
As we discuss more explicitly below, a direct expansion of electron self-energy in powers of the Coulomb interaction
is never possible in a 2DES because of the long-range of the Coulomb interaction.  Our results for the C2DES are based on the 
random phase approximation (RPA) in which the self-energy is expanded to first order in the dynamically screened 
Coulomb interaction $W$ (setting $\hbar=1$): 
\begin{equation}\label{eq:sigma_rpa}
\label{eq:RPAse}
\Sigma_s({\bf k},i\omega_n)=-\frac{1}{\beta}\sum_{s'}
\int \frac{d^2{\bf q}}{(2\pi)^2}\sum_{m=-\infty}^{+\infty}
W({\bf q},i\Omega_m)\left[\frac{1+s s'\cos{(\theta_{{\bf k},{\bf k}+{\bf q}})}}{2}\right]G^{(0)}_{s'}({\bf k}+{\bf q},i\omega_n+i\Omega_m)\,,
\end{equation}
where $s=+$ for electron-doped systems and $s=-$ for 
hole-doped systems, $\beta=1/(k_{\rm B} T)$, 
\begin{equation}\label{eq:ex+corr}
W({\bf q},i\Omega)=v_q+v^2_q\chi_{\rho\rho}({\bf q},i\Omega)\,,
\end{equation}
\begin{equation}\label{eq:chi_RPA}
\chi_{\rho\rho}({\bf q},i\Omega) =
\frac{\chi^{(0)}({\bf q},i\Omega)}{1-v_q\chi^{(0)}({\bf q},i\Omega)} \equiv \frac{\chi^{(0)}({\bf q},i\Omega)}{\varepsilon({\bf q},i\Omega)}
\end{equation} 
is the RPA density-density response function, $\chi^{(0)}$ is its non-interacting limit ({\it 21--24}), 
and $\varepsilon({\bf q},i\Omega)$ is the RPA dielectric function. For definiteness, we limit our discussion to
an electron-doped system with positive chemical potential $\mu$: the Fermi liquid properties at negative doping are 
identical because of the C2DES model's particle-hole symmetry (see also Fig.~2). 

In Eq.~(\ref{eq:sigma_rpa}) $\omega_n=(2n+1)\pi/\beta$ is a fermionic Matsubara frequency, the sum runs over all the bosonic Matsubara frequencies $\Omega_m=2m\pi/\beta$ while in Eqs.~(\ref{eq:ex+corr}) and~(\ref{eq:chi_RPA}), $v_q$ is the bare unscreened Coulomb interaction
in 2D, $v_q = 2 \pi e^2/(\epsilon q)$ where $\epsilon$ is an effective dielectric constant. 
The first and second terms in Eq.~(\ref{eq:ex+corr}) are responsible respectively for the exchange interaction with the 
occupied Fermi sea (including the negative energy component), 
and for the interaction with particle-hole and collective virtual fluctuations.  The factor 
in square brackets in Eq.~(\ref{eq:RPAse}), which depends on the angle $\theta_{{\bf k},{\bf k}+{\bf q}}$ between ${\bf k}$ and ${\bf k}+{\bf q}$, captures
the dependence of Coulomb scattering on the relative chirality $s s'$ of the interacting electrons. 
The Green's function $ G^{(0)}_{s}({\bf k},i\omega) = 1/[i\omega - \xi_s({\bf k})]$ describes the free propogation of 
states with wavevector ${\bf k}$, Dirac energy $ \xi_s({\bf k})=sv k-\mu$ (relative to the chemical potential) and chirality $s=\pm$. 
After continuation from imaginary to real frequencies, $i \omega \to 
\omega + i \eta$, the quasi-particle weight factor 
$Z$ and the renormalized Fermi velocity can be expressed ({\it 20}) in terms of the wavevector and 
frequency derivatives of the retarted self-energy $\Sigma^{\rm ret}_+({\bf k},\omega)$ evaluated at the Fermi surface ($k=k_{\rm F}$) and 
at the quasiparticle pole $\omega=\xi_+({\bf k})$:
\begin{equation}\label{eq:Z_def}
Z= \frac{1}{1-\left.\partial_{\omega} \Re e \Sigma^{\rm ret}_+({\bf k},\omega)\right|_{k=k_{\rm F},\omega=0}}\,,
\end{equation}
and
\begin{equation}\label{eq:v_star_dyson}
\frac{v^\star}{v}
=\frac{\displaystyle 1+(v)^{-1}\left.\partial_k \Re e \Sigma^{\rm ret}_+({\bf k},\omega)\right|_{k=k_{\rm F},\omega=0}}{1-\left.
\partial_{\omega} \Re e \Sigma^{\rm ret}_+({\bf k},\omega)\right|_{k=k_{\rm F},\omega=0}}\,.
\end{equation}
Following some standard manipulations ({\it 20)} the self-energy can be expressed in 
a form convenient for numerical evaluation, as the sum of a contribution from the interaction of quasiparticles
at the Fermi energy, the {\em residue} contribution $\Sigma^{\rm res}$, and a contribution from interactions with quasiparticles
far from the Fermi energy and via both exchange and virtual fluctuations, the {\em line} contribution $\Sigma^{\rm line}$.
In the zero-temperature limit 
\begin{eqnarray}\label{eq:residue_t_0}
\Sigma^{\rm res}_+({\bf k},\omega)&=&\sum_{s'}\int \frac{d^2 {\bf q}}{(2\pi)^2}
\frac{v_q}{\varepsilon({\bf q},\omega-\xi_{s'}({\bf k}+{\bf q}))}\left[\frac{1+s'\cos{(\theta_{{\bf k},{\bf k}+{\bf q}})}}{2}\right]\nonumber\\
&\times&\left[\Theta(\omega-\xi_{s'}({\bf k}+{\bf q}))-\Theta(-\xi_{s'}({\bf k}+{\bf q}))\right]
\end{eqnarray}
and
\begin{equation}\label{eq:line_t_0_better}
\Sigma^{\rm line}_+({\bf k},\omega)=-\sum_{s'}
\int \frac{d^2 {\bf q}}{(2\pi)^2}v_q\left[\frac{1+s'\cos{(\theta_{{\bf k},{\bf k}+{\bf q}})}}{2}\right]\int_{-\infty}^{+\infty}\frac{d\Omega}{2\pi}
\frac{1}{\varepsilon({\bf q},i\Omega)}\frac{\omega-\xi_{s'}({\bf k}+{\bf q})}{[\omega-\xi_{s'}({\bf k}+{\bf q})]^2+\Omega^2}\,.
\end{equation}
Note that at the Fermi energy $\partial_{k} \Sigma^{\rm res}_+({\bf k},\omega)$ vanishes, 
and $\partial_{\omega} \Sigma^{\rm res}_+({\bf k},\omega)$ involves an integral over interactions on the 
Fermi surface that are statically screened.  These expressions differ from the corresponding 2DES expressions 
because of the relative chirality dependence of the Coulomb matrix elements, because of the 
linear dispersion of the bare quasiparticle energies, and most importantly because of the 
fast short-wavelength density fluctuations produced by the interband contribution to 
$\chi^{(0)}({\bf q},i\Omega)$ illustrated in Fig.~3.

Our results for $Z$ and $v^\star/v$ are summarized in Fig.~4 as a function of the C2DES dimensionless coupling constant (restoring $\hbar$)
\begin{equation}
\label{eq:coupling}
f \equiv \nu\frac{2\pi e^2}{\epsilon k_{\rm F}}=g \, \frac{e^2}{\epsilon v \hbar}\,.
\end{equation}
The appropriate value of $f$ for a particular graphene sheet 
is dependent on its dielectric environment; for graphene on ${\rm SiO}_2$ $f \sim 2$.  
As we discuss at greater length below, graphene's Fermi liquid properties depend only weakly 
on the carrier density which is expressed in these figures in terms of the cut-off parameter 
$\Lambda$.  The trends exhibited in Fig.~4 can be understood by considering the limits 
of small $f$ and the limit of large $q$ at all values of $f$.  
In the former limit screening is weak except at extremely small $q$.
In $\partial_{\omega} \Sigma^{\rm res}_+({\bf k},\omega)$, for example, the integral over $q$ diverges
logarithmically at small $q$ when $\varepsilon({\bf q},\omega=0)$ is set equal to one, {\em i.e.} when 
screening is neglected.  Screening cuts off this logarithmic divergence at a wavevector proportional to 
$f$ so that $\partial_{\omega} \Sigma^{\rm res}_+({\bf k},\omega)$ has a contribution proportional to $f \ln(f)$ 
at small $f$.  Because $\varepsilon({\bf q},\omega=0)$ happens to be independent of $q$ for transitions between
Fermi surface points, it is possible to evaluate $\partial_{\omega} \Sigma^{\rm res}_+({\bf k},\omega)$
analytically.  We find that 
\begin{eqnarray}\label{eq:partial_omega_residue_final_4}
\left.\frac{\partial}{\partial \omega}\Re e \Sigma^{\rm res}_+({\bf k},\omega)\right|_{k=k_{\rm F}, \omega=0}&=&\frac{f}{2\pi g}
\left[\sqrt{4-f^2}\ln{\left(\frac{2+\sqrt{4-f^2}}{f}\right)}-\frac{1}{2}(4-f\pi)\right]\,.
\end{eqnarray}
Similar small $q$ $f \ln (f)$ contributions appear in the other elements which contribute to 
$Z$ and $v^\star$.  All this behavior is very familiar from the case of the normal 2DES; the new differences
present in the chiral C2DEG are ones of detail.  At large $q$, on the other hand, interband charge fluctuations 
dominate  $\varepsilon({\bf q},\omega)-1$, which approaches 
its simple undoped system form. It becomes especially clear when $\omega$ is expressed in units of $vq$ that the 
typical value of $\varepsilon({\bf q},\omega)$ at large $q$ is $\sim 1$ with a non-trivial dependence on $f$.
The $q$ integrals all vary as $q^{-1}$, requiring that the C2DES model be accompanied by an ultraviolet 
cut-off which for the case of graphene should be~({\it 24}) $q_{\rm c} \sim 1/a$ where $a$ is the 
graphene lattice constant.  Since the crossover between intraband and interband screening occurs for 
$q \sim k_{\rm F}$, it follows that both $\partial_k \Sigma^{\rm line}$ and 
$\partial_\omega \Sigma^{\rm line}$ have contributions that are analytic in $f$ and vary as 
$\ln(\Lambda)$ where $\Lambda=q_{\rm c}/k_{\rm F}$ when $\Lambda$ is large.  To leading order in $\ln(\Lambda)$ we find that 
\begin{equation}\label{eq:Z_ln}
Z^{-1}- 1 = \frac{f \lambda(f)}{6g}\ln{(\Lambda)}
\end{equation}
and that~({\it 25}) 
\begin{equation}\label{eq:v_star_osa_ln}
\frac{v^\star}{v}- 1 = \frac{f[1-f\xi(f)]}{4g}\ln{(\Lambda)}
\end{equation}
where 
\begin{equation}
\lambda(f)=\frac{48}{\pi}\int_{0}^{+\infty}dx
\frac{1}{8\sqrt{1+x^2}+f\pi}\frac{x^2-1}{(1+x^2)^{3/2}}\,
\end{equation}
and 
\begin{equation}\label{eq:xi}
\xi(f)=4\int_0^{+\infty}dx
\frac{1}{8\sqrt{1+x^2}+f\pi}\frac{1}{(1+x^2)^{2}}\,.
\end{equation}
Note that $\lambda(f)=f/4-3\pi^2 f^2/256+...$ and $\xi(f)=1/3-3\pi^2 f/256+...$ are analytic functions of $f$ 
because interband polarization screening does not essentially alter the Coulomb interaction at large $q$.

The asymptotic expressions (\ref{eq:Z_ln}) and (\ref{eq:v_star_osa_ln}) approximately capture the 
contribution to the corresponding Fermi liquid parameters from interactions over the 
wavevector range from $\sim k_{\rm F}$ to $\sim q_{\rm c}$. 
As the density decreases and $k_{\rm F}\to 0$ this contribution dominates.
The Fermi wavelength then acts like a cut-off on the renormalization group 
flows which appear in the theory ({\it 26}) of interaction effects in undoped 
graphene.  The fact that the velocity increases in this regime can be understood qualitatively using Hartree-Fock theory~({\it 24}), 
which is accurate at small $f$ when $\Lambda$ is large.  In Hartree-Fock theory the enhanced 
velocity is due to the reduced exchange energy in a right-handed band when the negative energy sea is left-handed. 
In Fig.~4 we have also shown cut-off and coupling constant dependence of the antisymmetric Landau parameter $F^{\rm a}_0$, 
which is defined in terms for the antisymmetric Landau interaction function $f_{\rm a}(\cos(\varphi))$ as~({\it 20})
\begin{equation}\label{eq:f_ell}
F^{\rm a}_\ell=\nu^\star\int_0^{2\pi}\frac{d\varphi}{2\pi}f_{\rm a}(\cos(\varphi))\cos{(\ell\varphi)}\,,
\end{equation}
where $\nu^\star=g k_{\rm F}/(2\pi v^\star)$ is the density-of-states of the interacting system at the Fermi surface 
[$f_{\rm a}(\cos(\varphi))$ is obtained from the Fermi surface dependence of the self-energy ({\it 27})]. 
Physically, $F^{\rm a}_0$ determines the spin susceptibility $\chi_{\rm S}=(v/v^\star)/(1+F^{\rm a}_0)$.  From our results in panels (b) and (c) of Fig.~4 we predict a rather large suppression of the spin susceptibility which could be measured in weak-field Shubnikov-de Haas magnetotransport experiments using a tilted magnetic field to distinguish spin and orbital response~({\it 14}). 

Our findings have important implications for density-functional-theory (DFT) 
and tight-binding modeling of ribbons ({\it 28--30}), quantum dots ({\it 31, 32}), and other 
nanostructures made from graphene.  Because of the pseudo-chiral properties of bulk quasiparticles, 
states tend to have a lower energy when they have the majority chirality.  This interaction effect 
depends specifically on intersite coherence and is 
completely missing in the local-density-approximation and in other approximations for exchange 
and correlation potentials commonly used in DFT.  The accuracy of graphene nanostructure 
electronic structure calculations would be improved if they used exchange-correlation potentials based on the
properties of the C2DES rather than on the properties of the ordinary 2DES.   

This work has been supported by the Welch Foundation, by the Natural Sciences and Engineering Research Council of Canada,
by the Department of Energy under grant DE-FG03-02ER45958, and by the National Science Foundation under grant DMR-0606489.

\vspace{0.6cm}
\noindent

\subsection*{References and Notes}
\begin{itemize}
\item[1.] K.S. Novoselov {\em et al.}, {\it Science} {\bf 306}, 666 (2004).
\item[2.] K.S. Novoselov {\em et al.}, {\it Nature} {\bf 438}, 197 (2005).
\item[3.] Y.B. Zhang {\em et al.}, {\it Nature} {\bf 438}, 201 (2005).
\item[4.] J.C. Slonczewski and P.R. Weiss, {\it Phys. Rev.} {\bf 109}, 272 (1958).
\item[5.] T. Ando, A.B. Fowler, and F. Stern, {\it Rev. Mod. Phys.} {\bf 54}, 437 (1982). 
\item[6.] K. von Klitzing, G. Dorda, and M. Pepper, {\it Phys. Rev. Lett.} {\bf 45}, 494 (1980).
\item[7.] D.C. Tsui, H.L. Stormer, and A.C. Gossard, {\it Phys. Rev. Lett.} {\bf 48}, 1559 (1982).
\item[8.] Y. Zhang {\em et al.}, {\it Phys. Rev. Lett.} {\bf 96}, 136806 (2006). 
\item[9.] K.S. Novoselov {\em et al.}, {\it Nature Physics} {\bf 2}, 177 (2006). 
\item[10.] K. Nomura and A.H. MacDonald, {\it Phys. Rev. Lett.} {\bf 96}, 256602 (2006).  
\item[11.] D. Pines and P. Nozi\'eres, {\em The Theory of Quantum Liquids} (Addison-Wesley, Menlo Park, 1966). 
\item[12.] R. Shankar, {\it Rev. Mod. Phys.} {\bf 66}, 129 (1994).  
\item[13.] E. Tutuc, S. Melinte, and M. Shayegan, {\it Phys. Rev. Lett.} {\bf 88}, 036805 (2002).
\item[14.] J. Zhu, H. L. Stormer, L. N. Pfeiffer, K. W. Baldwin, and K. W. West, {\it Phys. Rev. Lett.} {\bf 90}, 056805 (2003). 
\item[15.] Y.W. Tan, J. Zhu, H.L. Stormer, L.N. Pfeiffer, K.W. Baldwin, and K.W. West, {\it Phys. Rev. B} {\bf 73} 045334 (2006). 
\item[16.] S. De Palo, M. Botti, S. Moroni, and G. Senatore, {\it Phys. Rev. Lett.} {\bf 94}, 226405 (2005).
\item[17.] R. Asgari, B. Davoudi, M. Polini, G.F. Giuliani, M.P. Tosi, and G. Vignale, {\it Phys. Rev. B} {\bf 71}, 045323 (2005).
\item[18.] Y. Zhang and S. Das Sarma, {\it Phys. Rev. B} {\bf 72}, 075308 (2005).
\item[19.] J.P. Eisenstein, L.N. Pfeiffer, and K.W. West,  {\it Phys. Rev. B} {\bf 50}, 1760 (1994).  
\item[20.] For a thorough discussion of the random phase approximation applied to ordinary electron gases see 
G.F. Giuliani and G. Vignale, {\it Quantum Theory of the Electron Liquid} (Cambridge University Press, Cambridge, 2005).
The random phase approximation for the Fermi liquid properties of graphene has also
been discussed recently by S. Das Sarma, E.H. Hwang, and Wang-Kong Tse,
Phys. Rev. B {\bf 75}, 121406(R) (2007).  The results presented here differ in several important 
respects, but both works agree that doped graphene is a Fermi liquid. 
\item[21.] K.W.-K. Shung, {\it Phys. Rev. B} {\bf 34}, 979 (1986).
\item[22.] B. Wunsch, T. Stauber, F. Sols, and F. Guinea,  {\it New J. Phys.} {\bf 8}, 318 (2006).
\item[23.] E.H. Hwang and S. Das Sarma, cond-mat/0610561.
\item[24.] Y. Barlas, T. Pereg-Barnea, M. Polini, R. Asgari, and A.H. MacDonald, cond-mat/0701257.
\item[25.] Eq.~(\ref{eq:v_star_osa_ln}) is obtained after performing a weak-coupling expansion on Eq.~(\ref{eq:v_star_dyson}).
\item[26.] J. Gonz\'alez, F. Guinea, and M.A.H. Vozmediano, {\it Phys. Rev. B} {\bf 59}, 2474(R) (1999).
\item[27.] T.M. Rice, {\it Ann. Phys.} {\bf 31}, 100 (1965). 
\item[28.] V. Barone, O. Hod, and G.W. Scuseria, {\it Nano Lett.} {\bf 6}, 2748 (2006). 
\item[29.] F. Mun\~oz-Rojas, D. Jacob, J. Fern\'andez-Rossier, and J.J. Palacios, {\it Phys. Rev. B} {\bf 74}, 195417 (2006).
\item[30.] Y.W. Son, M.L. Cohen, and S.G. Louie, {\it Nature} {\bf 44}, 347 (2006). 
\item[31.] P.G. Silvestrov and K.B. Efetov, {\it Phys. Rev. Lett.} {\bf 98}, 016802 (2007). 
\item[32.] A. De Martino, L. Dell'Anna, and R. Egger, {\it Phys. Rev. Lett.} {\bf 98}, 066802 (2007).
\end{itemize}

\newpage

\begin{figure}[t]
\begin{center}
\includegraphics[width=0.5\linewidth]{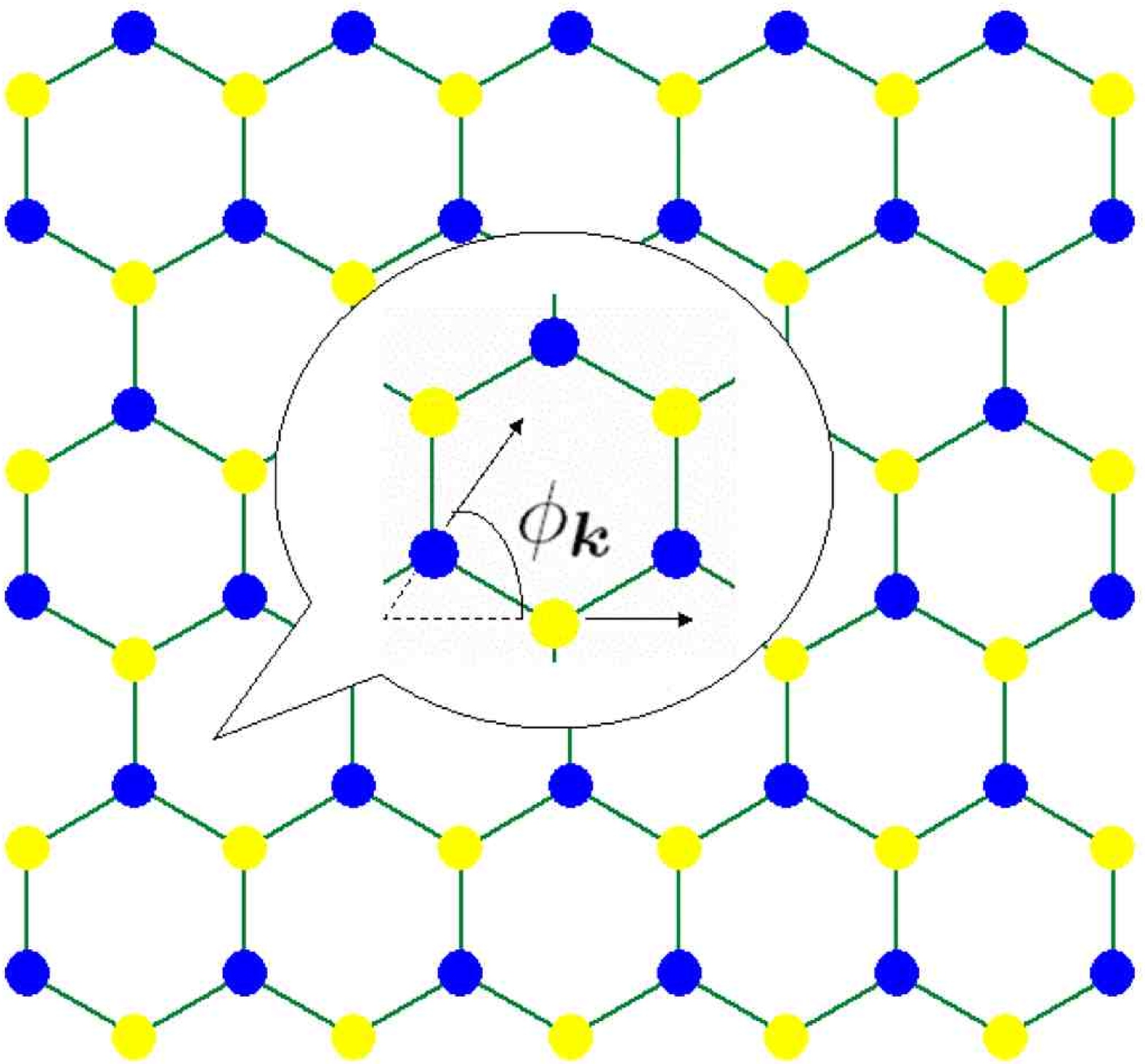}
\caption{Honeycomb lattice of a single layer graphite flake with one sublattice in yellow and 
the other sublattice in blue. In the continuum limit the sublattice degree of freedom may be regarded 
as a pseudospin.  When momentum ${\bf k}$ is measured away from the Dirac points at the $K$ and $K'$ Brillouin zone 
corners, band eigenstates have definite projection of pseudospin in the ${\bf k}$ direction,
{\em i.e.} definite pseudochirality. The angle $\phi_{\bf k}$ above denotes the momentum-dependent phase difference 
between wavefunction amplitudes on the two sublattices.  For spin-$1/2$ quantum particles this angle is the azimuthal orientation of a pseudospin 
coherent state in the equatorial plane.}
\label{fig:one}
\end{center}
\end{figure}

\begin{figure}[t]
\begin{center}
\includegraphics[width=0.5\linewidth]{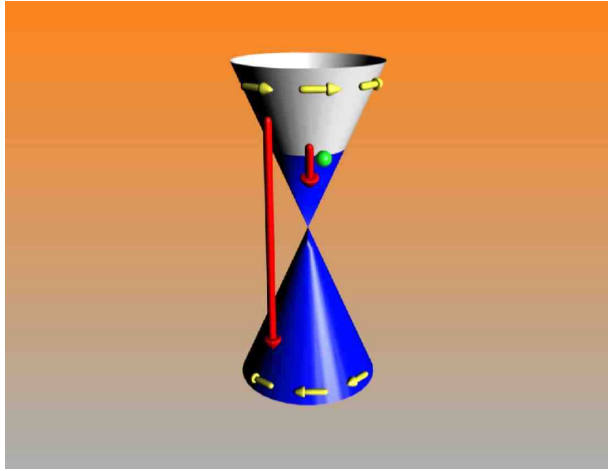}
\caption{In a weakly doped material, graphene's energy bands can be described by a massless Dirac equation in which the role of spin is played by pseudospin.  Like an ordinary 2DES, doped graphene has a circular Fermi surface.  The Fermi liquid properties of 
graphene are a consequence of both exchange interactions between quasiparticles near the Fermi surface and states in the positive and negative energy Fermi seas and of interactions with both intra-band (short red vertical arrow) and inter-band (long red vertical arrow) virtual fluctuations of the electronic system. The yellow arrows in this figure indicate the pseudospin chirality of band eigenstates. Because of the difference in chirality between positive and negative energy bands, the velocity of graphene quasiparticles is enhanced by inter-band exchange interactions, tending to protect the system from magnetic and other instabilities, and reducing both charge and spin response functions.}
\label{fig:two}
\end{center}
\end{figure}

\begin{figure}[t]
\begin{center}
\includegraphics[width=0.8\linewidth]{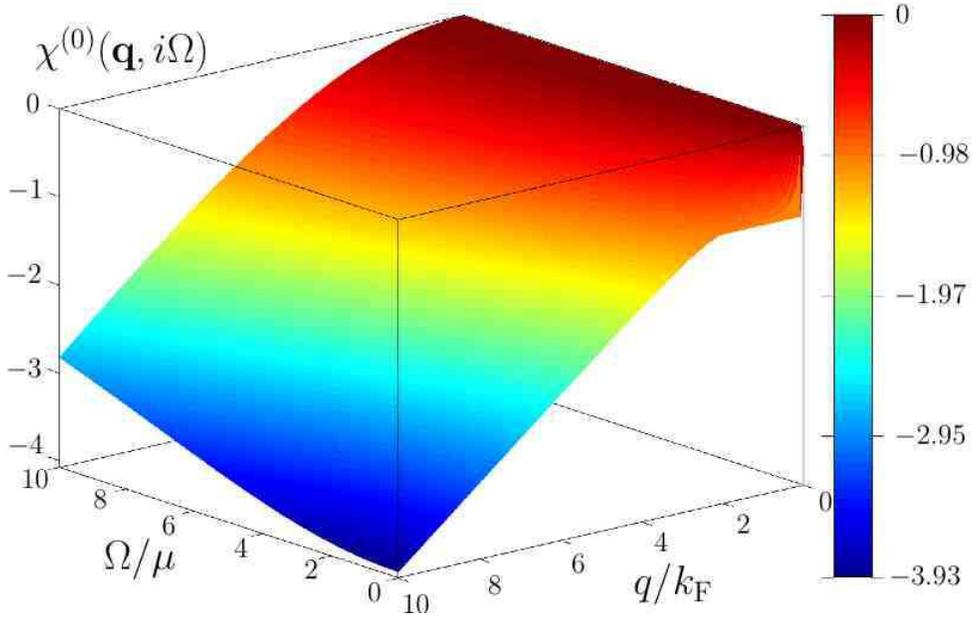}
\caption{``Lindhard" function $\chi^{(0)}({\bf q},i\Omega)$ of a C2DES, in units of the non-interacting density-of-states at the Fermi surface 
$\nu=g k_{\rm F}/(2\pi v)$, as a function of $q/k_{\rm F}$ and $\Omega/\mu$ on the imaginary frequency axis. 
$k_{\rm F}=(4\pi n/g)^{1/2}$ is the Fermi wavenumber, $\mu=v k_{\rm F}$ the Fermi energy, $n$ the electron density and 
the flavor multiplicity $g = g_{\rm s} g_{\rm v} = 4$ for graphene because of its two-fold valley degeneracy. 
Because of interband fluctuations $\chi^{(0)}$ diverges linearly with $q$ for $q\to \infty$ and 
decays only like $\Omega^{-1}$ for $\Omega \to \infty$ in the C2DES, in contrast to the $q^{-2}$ and $\Omega^{-2}$ behaviors of the ordinary 2DES.  
In the static $\Omega=0$ limit 
$\chi^{(0)}({\bf q},0)=-\nu$ for all $q\leq 2k_{\rm F}$ 
for both chiral and ordinary 2DESs. \label{fig:three}}
\end{center}
\end{figure}

\begin{figure}[t]
\begin{center}
\tabcolsep=0cm
\begin{tabular}{ccc}
\includegraphics[width=0.33\linewidth]{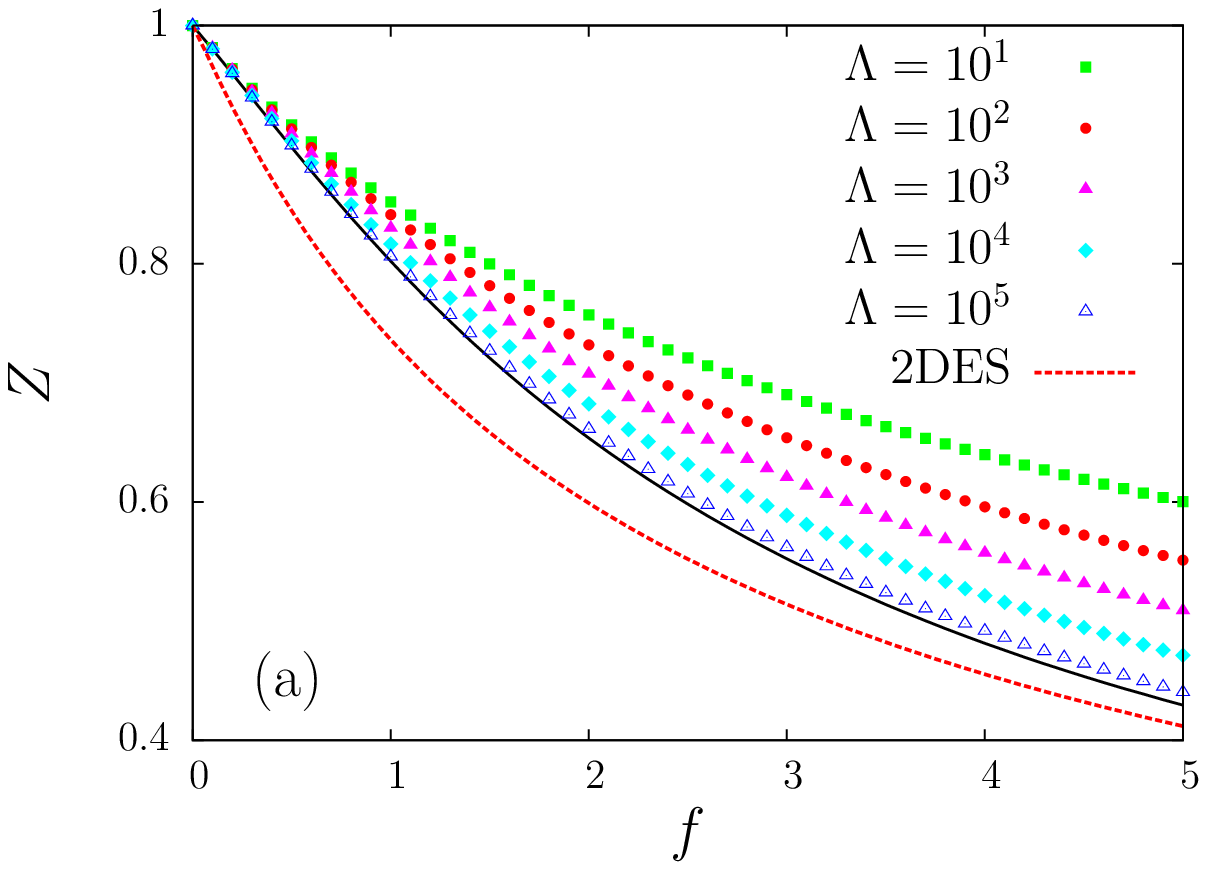}&
\includegraphics[width=0.34\linewidth]{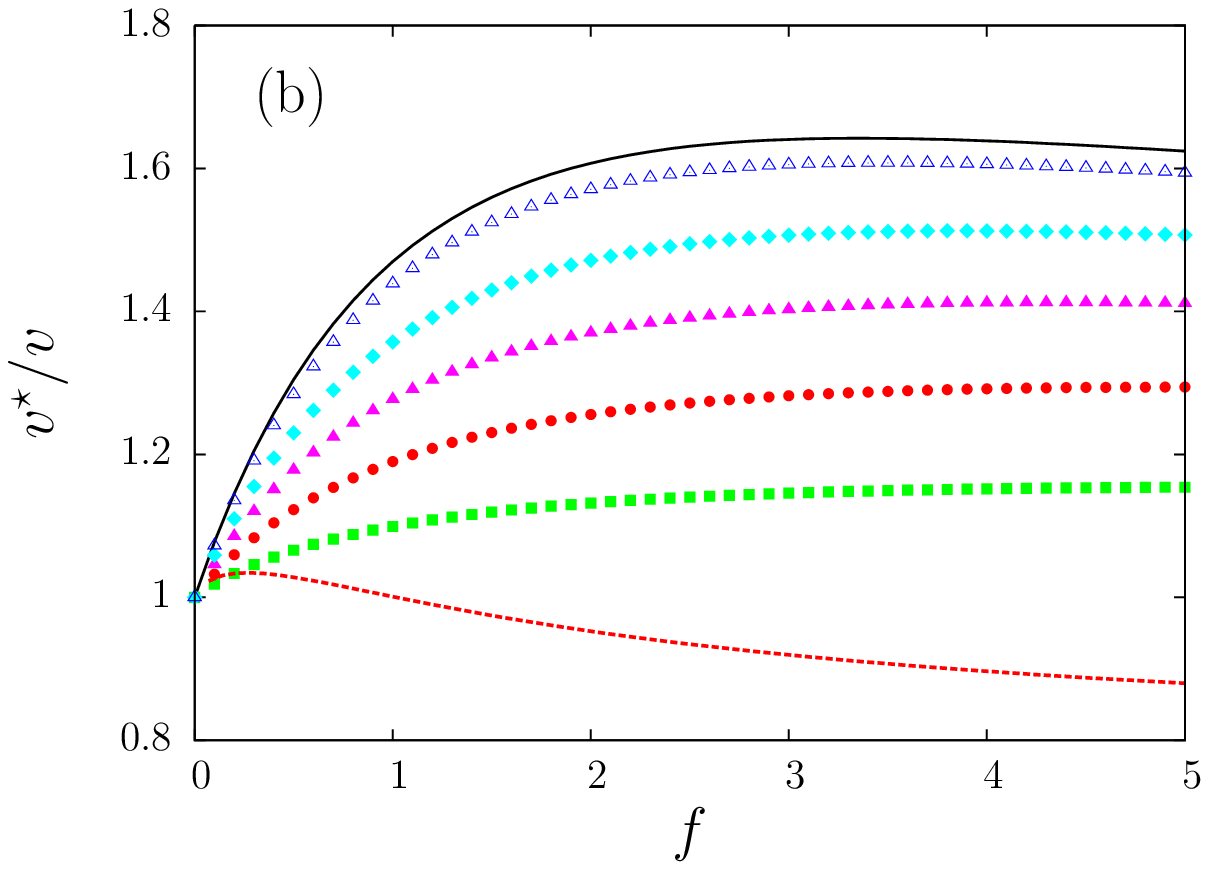}&
\includegraphics[width=0.33\linewidth]{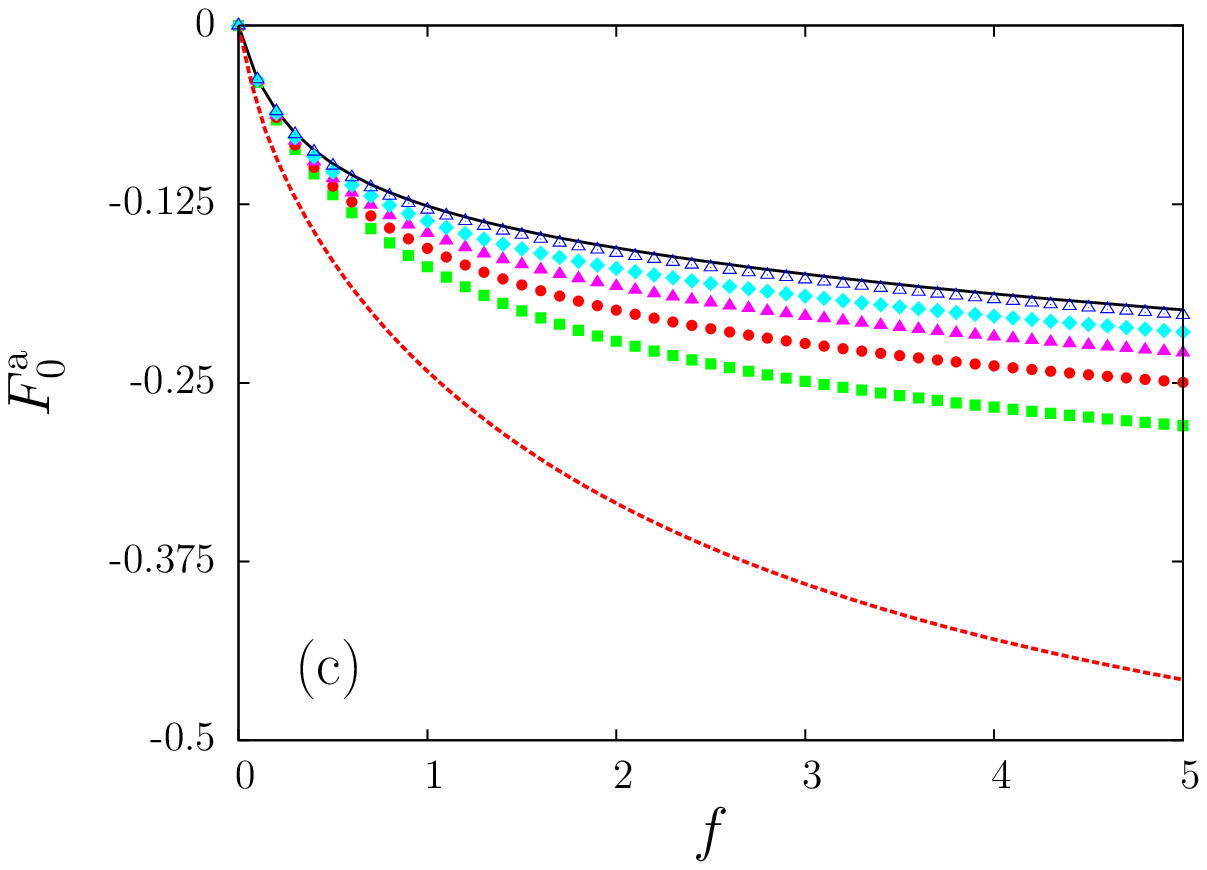}
\end{tabular}
\end{center} 
\caption{Density and coupling constant $f$ dependence of some C2DES Fermi-liquid parameters.  The 
density is specified by $\Lambda \equiv q_{\rm c}/k_{\rm F}$.
The density range studied most extensively in experiment, $n\sim 10^{11}~{\rm cm}^{-2}$ to $n\sim 10^{13}~{\rm cm}^{-2}$, 
corresponds to $\Lambda=100$ to
$\Lambda=10$.  In all panels the black solid line corresponds to the highest value of the cut-off parameter we have 
considered, $\Lambda=2.7\times 10^5$.  The red dashed line illustrates the RPA Fermi-liquid parameters of an ordinary non-chiral 2DES with parabolic bands. In this case the $f=\sqrt{2}~r_s$ [see Eq.~(\ref{eq:coupling})], where $r_s=(\pi n a^2_{\rm B})^{-1/2}$ is the usual Wigner-Seitz density parameter and $a_{\rm B}=\epsilon\hbar^2/(m_{\rm b} e^2)$ the effective Bohr radius. From the left the three panels show: (a) the quasiparticle renormalization factor $Z$ evaluated from Eq.~(\ref{eq:Z_def}); (b) the 
velocity renormalization factor evaluated from Eq.~(\ref{eq:v_star_dyson}); and (c) the $\ell=0$ dimensionless Landau parameter $F^{\rm a}_0$ which characterizes spin-dependent quasiparticle interactions. The color coding for $\Lambda$ is the same in 
all panels.}
\end{figure}

\clearpage

\end{document}